\documentclass[aps,prl,twocolumn,groupedaddress]{revtex4-1}
\usepackage{graphicx,epsfig,dcolumn,afterpage}
\usepackage{colortbl}
\usepackage[latin1]{inputenc} 
\usepackage{amsmath}
\usepackage{amssymb}  

\usepackage{color}
\usepackage{colortbl}

\newcommand{\TV}[1]{}  

\newcommand{\xten}[1]{{}\times 10^{#1}}

\newcommand{\eV}{\,\text{eV}}

\newcommand{\Wcm}{\,\text{W}/\text{cm}^2}

\newcommand{\fs}{\,\text{fs}}

\begin{document}

\title{Controlling ultrafast currents by the non-linear photogalvanic effect}

\author{Georg Wachter$^1$}
\email{georg.wachter@tuwien.ac.at}
\author{Shunsuke A. Sato$^2$}
\author{Christoph Lemell$^1$}
\author{Xiao-Min Tong$^{2,3}$}
\author{Kazuhiro Yabana$^{2,3}$}\
\author{Joachim Burgd\"orfer$^1$}

\affiliation{$^1$Institute for Theoretical Physics, Vienna University of Technology, 1040 Vienna, Austria, EU}
\affiliation{$^2$Graduate School of Pure and Applied Sciences, University of Tsukuba, Tsukuba 305-8571, Japan}
\affiliation{$^3$Center for Computational Sciences, University of Tsukuba, Tsukuba 305-8577, Japan}

\date{\today}

\begin{abstract}
We theoretically investigate the effect of broken inversion symmetry on the generation and control of ultrafast currents in a transparent dielectric (SiO$_2$) by strong femto-second optical laser pulses. 
Ab-initio simulations based on time-dependent density functional theory predict ultrafast DC currents 
that can be viewed as a non-linear photogalvanic effect. 
Most surprisingly, the direction of the current undergoes a sudden reversal above a critical threshold value of laser intensity $I_c \sim 3.8 \xten{13} \Wcm$. 
We trace this switching to the transition from non-linear polarization currents to the tunneling excitation regime. 
We demonstrate control of the ultrafast currents by the time delay between two laser pulses.
We find the ultrafast current control by the non-linear photogalvanic effect to be remarkably robust and insensitive to laser-pulse shape and carrier-envelope phase. 
\end{abstract}



\maketitle

In the last decade, ultrafast few-cycle laser pulses with well-defined carrier-envelope phase (CEP) have become available providing novel opportunities to explore the ultrafast and non-linear response matter to strong optical fields. 
The study of the induced electronic motion and of the highly non-linear optical response have focussed on rare gas atoms \cite{krausz_attosecond_2009}, molecules \cite{scrinzi_attosecond_2006} and, more recently, on nanostructures, surfaces and bulk matter \cite{cavalieri_attosecond_2007,gertsvolf_orientation-dependent_2008,*gertsvolf_demonstration_2010}.
The driven electron dynamics can be monitored through optical signals \cite{mitrofanov_optical_2011,ghimire_observation_2011,schultze_controlling_2013,schultze_attosecond_2014,schubert_sub-cycle_2014} and through emitted  electrons \cite{lemell_electron_2003,dombi_direct_2004,kruger_attosecond_2011,wachter_electron_2012-1,zherebtsov_controlled_2011,neppl_direct_2015}. 
Very recently, Schiffrin \emph{et al.} \cite{schiffrin_optical-field-induced_2013} have demonstrated directed electron currents generated inside transparent dielectrics by carefully tailored laser pulses. In turn, the ultrafast response can characterize the impinging laser field \cite{paasch-colberg_solid-state_2014}. Currently, avenues are explored to exploit such ultrafast modulation of electric currents for petahertz-scale signal processing \cite{krausz_attosecond_2014} enabled by the short intrinsic time scale of the electron motion ($\sim 1 \fs$), orders of magnitude faster than semiconductor electronics. 

In this letter, we explore a novel channel for the ultrafast electronic response that is unique to dielectrics with a non-centrosymmetric crystallographic structure: the generation of DC currents induced by strong optical laser pulses. 
Fully three-dimensional \emph{ab-initio} simulations based on time-dependent density functional theory (TD-DFT) predict the generation of strongly non-linear currents in $\alpha$-quartz that are, in contrast to previously observed currents \cite{schiffrin_optical-field-induced_2013,paasch-colberg_solid-state_2014}, \emph{independent} of the details of the laser pulse shape. The direction of the currents is found sensitive to the instantaneous laser intensity. Analysis of the spatio-temporal charge dynamics on the atomic length and time scale allow us to link this to the transition from non-linear polarization currents to directional tunneling excitation, the latter being highly sensitive to the alignment between the laser polarization and the chemical bonds in the crystal. We demonstrate that this transition may be investigated in a pump-probe setup leaving its marks as a change of the direction of the current as a function of the pump-probe delay. 

Theoretical exploration of ultrafast processes in solids faces the challenge to tackle the time-dependent many-body problem. Time-dependent density functional theory has emerged as a versatile tool allowing for an \emph{ab-initio} description of a variety of strong field processes in the solid state \cite{wachter_electron_2012-1,otobe_first-principle_2012,schultze_attosecond_2014,neppl_direct_2015}. Here, we employ a real-space, real-time formulation of TDDFT \cite{yabana_time-dependent_1996,bertsch_real-space_2000,onida_electronic_2002,marques_time-dependent_2004,otobe_first-principles_2009,wachter_ab_2014} for the electronic dynamics induced by strong few-cycle laser pulses in $\alpha$-SiO$_2$ ($\alpha$-quartz). Briefly, we solve the time-dependent Kohn-Sham equations (atomic units are used unless stated otherwise)
\begin{equation}
  i \partial_t \psi_i(\mathbf{r},t) = H(\mathbf{r},t)\psi_i(\mathbf{r},t) \quad , 
\label{eq:tdks}
\end{equation}
where $i$ runs over the occupied Kohn-Sham orbitals $\psi_i$. The Hamiltonian 
\begin{equation} \label{eq:hamiltonian}
H(\mathbf{r},t) = 
\frac{1}{2} \left( -i\mathbf{\nabla} + \mathbf{A}(t) \right)^2 
+ \hat{V}_\mathrm{ion} 
+ \int d\mathbf{r'} \frac{ n(\mathbf{r'},t) }{ | \mathbf{r}-\mathbf{r'} | } + \hat{V}_\mathrm{XC}(\mathbf{r},t)
\end{equation}
describes the system under the influence of a homogenous time-dependent electric field $\mathbf F(t)$ of amplitude $\mathbf F_0$ along $\hat a$ with vector potential $\mathbf A(t) = - \int_{-\infty}^t \mathbf F(t') dt' $ in the velocity gauge and in the transverse geometry \cite{yabana_time-dependent_2012} allowing to treat the bulk polarization response of the infinitely extended system along the polarization direction. The periodic lattice potential $\hat{V}_\mathrm{ion}$ is given by norm-conserving pseudopotentials of the Troullier-Martins form \cite{troullier_efficient_1991} representing the ionic cores (O(1s$^2$) and Si(1s$^2$2s$^2$2p$^6$)). The valence electron density is $n(\mathbf{r},t) = \sum_i |\psi_i(\mathbf{r},t)|^2$. 
For the exchange and correlation potential $\hat{V}_\mathrm{XC}$ we employ the adiabatic Tran-Blaha modified Becke-Johnson meta-GGA functional \cite{tran_accurate_2009,*koller_merits_2011,*koller_improving_2012}. It accurately reproduces the band gap $\Delta \sim 9$ eV for SiO$_2$ and yields good agreement with the experimental dielectric function over the spectral range of interest including at optical frequencies \cite{philipp_optical_1966}.
The time-dependent Kohn-Sham equations (Eq.~\ref{eq:tdks}) are solved on a Cartesian grid with discretization $\sim 0.25$ a.u. in laser polarization direction and $\sim 0.45$ a.u. perpendicular to the polarization direction in a cuboid cell of dimensions 9.28$\times$16.05$\times$10.21 a.u.$^3$ employing a nine-point stencil for the kinetic energy operator and a Bloch-momentum grid of $4^3$ $\mathbf k$-points. The time evolution is performed with a 4$^{\mathrm{th}}$-order Taylor approximation to the Hamiltonian with a time step of 0.02 a.u.~including a predictor-corrector step. 
The solution of Eqs.~\ref{eq:tdks} and \ref{eq:hamiltonian} allows to analyze the time and space dependent microscopic vectorial current density 
\begin{equation}
\mathbf j(\mathbf r, t) = |e| \sum_i \frac{1}{2} \left[ \psi^*_i(\mathbf r, t) \left( -i  \mathbf \nabla + \mathbf A(t) \right) \psi_i(\mathbf r, t) + \mathrm{c.c.} \right] \quad 
\end{equation}
as well as the mean current density $J(t)$ along the laser polarization direction $\mathbf F_0$, averaged over the unit cell of volume $\Omega$, 
\begin{equation}
J(t) = \frac{1}{\Omega} \int_\Omega d\mathbf r \,\, \mathbf j(\mathbf r,t) \cdot \mathbf{F_0}/|F_0|  \quad . 
\end{equation}
The polarization density $P(t)= \int_{-\infty}^t J(t') dt'$ \cite{resta_theory_2007} gives the charge 
density $D(t)$ transferred by the pulse. The total charge $Q$ will depend also on the details of the geometry of the laser focus and of the collection volume not explicitly treated in the following. 

First studies of the short-pulse induced current and charge transfer in polycrystalline SiO$_2$ \cite{schiffrin_optical-field-induced_2013,paasch-colberg_solid-state_2014,wachter_ab_2014} found a sinusoidal dependence on the carrier-envelope phase, $\phi_\mathrm{CE}$, of the few-cycle electromagnetic field $\mathbf A(t) \sim \mathbf A_0(t) \cos(\omega_L t + \phi_\mathrm{CE})$ with ($\cos^2$) envelope $\mathbf A_0(t)$ and $\omega_L$ the carrier frequency of the IR laser. Subcycle control and steering of electrons required exquisite control over the instantaneous electric field $\mathbf F(t)$. Here, we explore an alternate route towards steering, controlling, and switching ultrafast currents that does not rely on $\phi_\mathrm{CE}$ control of the instantaneous field but on the instantaneous intensity dependence of a direct (DC) current. 

\begin{figure}
\centerline{\epsfig{file=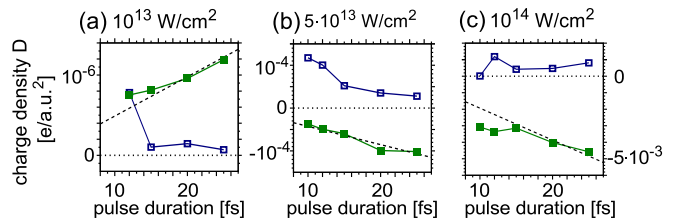,width=\columnwidth}}
\caption{(Color online) 
Pulse length dependence of transferred charge at intensities (a) $1\xten{13} \Wcm$, (b) $5\xten{13} \Wcm$, and (c) $1\xten{14} \Wcm$, each with $\cos^2$ pulse shape.  Carrier-envelope phase (CEP) dependent part $D_\mathrm{CEP}$ (blue open squares); CEP independent part $D_0$ (green full squares); dashed lines: linear slope going through the origin.
}
\label{fig1}
\end{figure}

Starting point is the observation that the total charge density $D(\tau_p)$ transferred at the conclusion of the pulse can be split into a CEP dependent part 
with amplitude $D_\mathrm{CEP}(\tau_p)$ and a residual part $D_0(\tau_p)$. 
$D_\mathrm{CEP}(\tau_p)$ tends to decrease with increasing pulse length while the magnitude of $D_0(\tau_p)$ increases with the pulse length (Fig.~\ref{fig1}). For pulse length exceeding a few optical cycles
, $D_0(\tau_p)$ is approximately proportional to the pulse duration and dominates the signal exceeding $D_\mathrm{CEP}$ by about one order of magnitude. 

\begin{figure}
\centerline{\epsfig{file=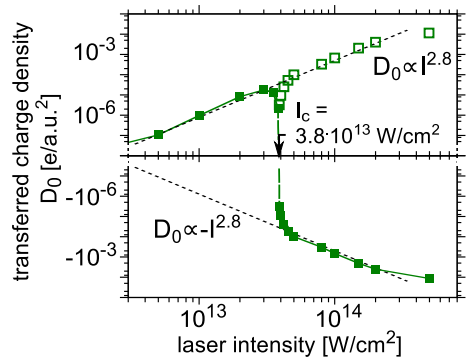,width=0.7\columnwidth}}
\caption{(Color online) 
Carrier-envelope phase \emph{independent} transferred charge density $D_0$ (green full squares) as function of laser intensity ($\cos^2$ pulse with full duration $\tau_p = 20 \fs$, $\hbar \omega_L = 1.7$ eV) and absolute value $|D_0|$ (empty boxes). Power law $|D_0| \propto I^{2.8}$ (dashed line). Electrons move along $+ \hat a$ for $I \le I_c = 3.8 \xten{13} \Wcm$ (upper panel) and along $- \hat a$ for $I \gtrsim I_c$ (broken line to guide the eye). 
}
\label{fig2}
\end{figure}

The charge transferred by the induced DC current, $D_0(\tau_p)$, features a strongly non-linear scaling with intensity $|D_0| \propto I^{2.8}$ or, equivalently, field strength $|D_0| \propto F_0^{5.6}$ (Fig.~\ref{fig2}). 
The origin of this highly non-linear response lies in the broken centrosymmetry of the SiO$_2$ crystal along the $\hat a$ direction. In general, generation of a directed flow of charge by a laser field requires a broken inversion symmetry. For few-cycle laser pulses with well-defined CEP, inversion symmetry is violated by a suitable choice of $\phi_\mathrm{CE}$. In the present case, it is no longer the temporal shape of the laser electric field but the electronic and crystallographic structure of matter the laser interacts with that causes ultrafast currents. This novel mechanism does not rely on delicate CEP control yet offers sub-cycle response and switching.

\begin{figure}
\centerline{\epsfig{file=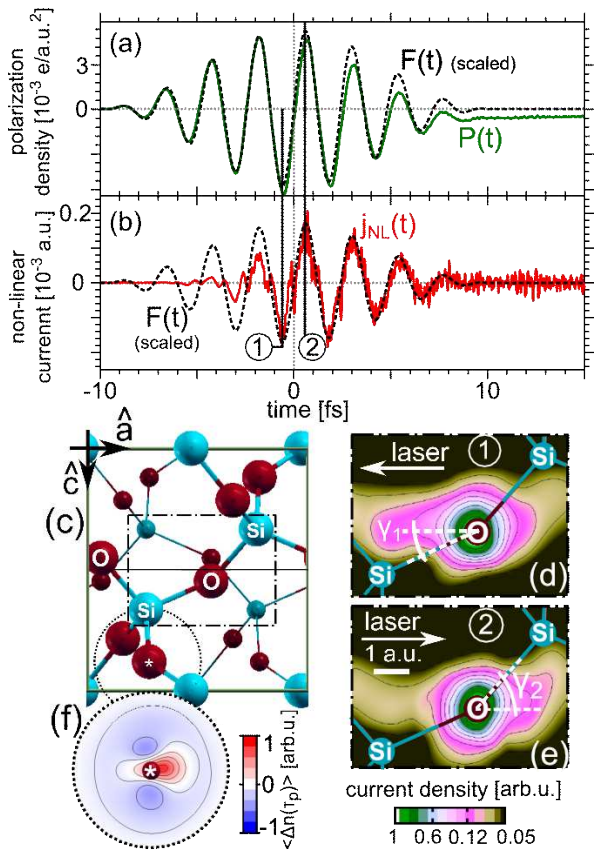,width=0.9 \columnwidth}}
\caption{(Color online) 
(a) Time-dependent polarization density $P(t)$ (solid green), laser field $F(t)$ (black dashed, scaled) for $\tau_p=20 \fs$, $\omega_L=1.7 \eV$, intensity $1 \xten{14} \Wcm$. 
(b) Time-dependent non-linear current $\Delta J_\mathrm{NL}$ (Eq.~\ref{eq:jnl}, red solid). 
(c) $\alpha$-quartz lattice with broken inversion symmetry ($\hat a \to -\hat a$). ``Larger'' atoms are closer to the reader than small atoms; labeled atoms form a helix along $\hat a$ direction. 
(d-e) Snapshot of the current density in an $\hat a$-$\hat c$ cut plane going through the central oxygen at a time \textcircled{1} (d), \textcircled{2} (e), dash-dotted rectangle in (c). Angles between electric field and O-Si bond are $\gamma_1 = 25.3$ deg, $\gamma_2 = 51.5$ deg. 
(f) Time-averaged density modulation after the laser pulse $\Delta n(\tau_p)$ in plane cut through the oxygen marked by a star in (c) for intensity $1 \xten{13} \Wcm$. 
\label{fig3}
}
\end{figure}

The appearance of a direct current in a homogeneous medium under illumination, independent of the CEP, and linearly increasing with pulse duration, can be viewed as a non-linear analogue to the well-known photogalvanic (PG) effect \citep{glass_highvoltage_1974,belinicher_photogalvanic_1980,sturman_photovoltaic_1992,fridkin_bulk_2001,glazov_high_2014} as first discussed qualitatively by Alon \cite{alon_bulk_2003}. 
Conventionally, the lowest order photogalvanic effect is described as 
\begin{equation} \label{eq:jpge}
  J_k^\mathrm{PG} = \beta_{kln} F_l F_n^* \,\, . 
\end{equation}
$J^\mathrm{PG}$ is quadratic in the electric field components and linear in the time-averaged laser intensity $I \propto F_l F_l^*$. For linearly polarized light, the photogalvanic tensor $\beta_{kln}$ associated with the two-wave mixing in the second-order susceptibility $\chi^{(2)}_{kln}(0; \omega, -\omega)$ is nonzero only in non-centrosymmetric crystals \cite{sipe_second-order_2000}. 
Microscopically, a variety of mechanisms may contribute to the PG effect such as asymmetric excitation, scattering, or recombination of electrons and electronic defects \cite{belinicher_photogalvanic_1980}. 
One important realization is the so-called ``shift current''  \citep{von_baltz_theory_1981,sipe_second-order_2000,young_first-principles_2012} due to the shift between the center of charge of the valence electrons and the excited electrons in the conduction band. This shift current has been predicted to be important in several semi-conductors \citep{sipe_second-order_2000,nastos_optical_2006,*nastos_optical_2010} and has been first experimentally verified for ferroelectrics \citep{young_first-principles_2012}. 

The present non-linear generalization of photogalvanic effects is obviously a signature of strong-field interaction with matter. This is underscored by the surprising observation of current reversal as a function of laser intensity (Fig.~\ref{fig2}). We find a critical value of current reversal at $I_c = 3.8 \xten{13} \Wcm$. 
Electrons move along the $+ \hat a$ direction for low intensities $I < I_c$ while they propagate along  $- \hat a$ direction for higher intensities $I \gtrsim I_c$. We elucidate the microscopic mechanism for this reversal by analysis of the spatio-temporal charge dynamics. 
At lower intensities $I < I_c$, the multi-photon driven non-linear polarization currents lead to a localized accumulation of charge in between the Si-O bonds as displayed in the time-averaged density fluctuations at the conclusion of the laser pulse  (Fig.~\ref{fig3}f). This implies the formation of an induced atomic-scale dipole around the oxygen atoms, i.e.~a displacement of the center of charge by vertical excitation, resembling the shift current mechanism of the standard photogalvanic effect but generalized to higher order reflected in the non-linear intensity scaling of $|D_0| \sim I^{2.8}$. 
For higher laser intensities $I>I_c$ the dominant charge transfer mechanism is excitation of the tilted conduction band by tunneling. 
Tunneling significantly depends on the local potential landscape in tunneling direction. We find tunneling is enhanced when the bond direction is aligned with the laser field as evidenced by a strongly asymmetric current density at times near the maxima of the electric field (Fig.~\ref{fig3}d,e). Tunneling excitation is more efficient along $- \hat a$ where the O-Si bond is more closely aligned with the laser field ($\gamma_1 = 25.3$ deg) while in $+ \hat a$ direction tunneling is suppressed because of the larger angle ($\gamma_2 = 51.5$ deg). 
The transition to tunneling excitation is therefore accompanied by a reversal of the charge transfer and current direction. As the tunneling rate scales exponentially with the peak intensity $\propto \exp(-1/\sqrt{I})$, the transition is quite abrupt suggesting its potential for femtosecond current switching. 

\begin{figure}
\centerline{\epsfig{file=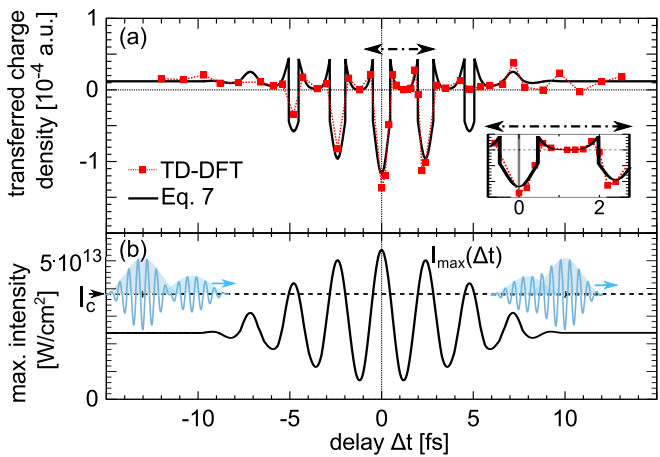,width=\columnwidth}}
\caption{(Color online) 
(a) Laser-induced transferred charge density as a function of the peak-peak delay ($I_1 = 2.4\xten{13}\Wcm$, $I_2 = 0.6\xten{13}\Wcm$, $\omega_L = 1.7 \eV$, $\tau_p = 20 \fs$). Red squares, dashed line: TD-DFT simulations; solid line: model Eq.~\ref{eq:model}. Inset: magnified data around $\Delta t \sim 0 $ to $ 2 \fs$ (dash-dotted arrow). 
(b) Maximum instantaneous laser intensity $I(t) = F(t)^2 (c / 8 \pi)$ for a given delay. Insets: pulse shape after superposition of pump and probe pulse $F(t)$ at $\Delta t = -13 \fs$ (weak pulse before strong pulse) and at $+10 \fs$. 
\label{fig4}
}
\end{figure}

For high intensities $I>I_c$, the charge transfer shows sub-cycle time structure. The time-dependence of the tunneling current can be conveniently analyzed by the non-linear response contribution $\Delta J_\mathrm{NL}(t)$ after subtracting the linear-response current scaled to the instantaneous field, 
\begin{equation} \label{eq:jnl}
  \Delta J_\mathrm{NL}(t) = J(t) - \frac{1}{2 \pi} \int_{-\infty}^{\infty} e^{-i \omega t} \sigma(\omega) F(\omega)
\end{equation}
with the conductivity $\sigma(\omega)$ determined for low intensity $I < I_c$ \cite{yabana_time-dependent_2012}. During the rise time of the pulse (Fig.~\ref{fig4}) $\Delta J_\mathrm{NL}$ is still $\approx 0$ as linear response prevails. 
However, once a field strength is reached sufficient for tunneling between neighboring atoms at around $-3 \fs$, the non-linear current shows strong spikes. While the linear response current is, to a good approximation, 90 degrees out of phase with the electric field and $P(t) = \int_{-\infty}^t dt' J(t')$ is in phase with $F(t)$, the current spikes are in phase with $F(t)$ as expected for tunneling excitation. 
At later times (from -1 fs), $\Delta J_\mathrm{NL}$ remains in phase with, but becomes proportional to the laser field, indicative of a conductor-like linear response $J(t) = \sigma_D F(t)$ with a Drude (free carrier-like) conductivity $\sigma_D$ for the tunneling-induced electron population in the conduction band.

The present analysis of the non-linear photogalvanic DC current suggests that the key control parameter is the instantaneous intensity $I(t)$ rather than the cycle averaged intensity in the conventional photogalvanic effect or the instantaneous value of the field $F(t)$ in the CE-phase controlled AC current. This sensitivity to $I(t)$ can be explored in a pump-probe setting, in which the instantaneous intensity can by manipulated by the delay between pump and probe pulses. The pump-probe delay may therefore serve as knob for fast charge transfer by the non-linear photogalvanic effect. To demonstrate this control we choose the intensity of both pump and probe pulse to be subcritical ($I_{1,2} < I_c$) with pump intensity $I_1= 2.4\xten{13} \Wcm$ and probe intensity $I_2= 0.6\xten{13} \Wcm$. However, the superimposed fields give rise to a maximum intensity of $I_\mathrm{max} = 2.25 I_1 = 5.4 \xten{13} \Wcm$ above $I_c$. The sign and amplitude of the induced current is controlled by the time delay between the laser pulses (Fig.~\ref{fig4}). For large positive and negative delays, the transferred charge saturates at the same positive value. In contrast, for near-zero delay $\Delta t = 0$ where the maximum intensity is attained, the DC current switches direction and the transferred charge becomes negative. Remarkably, during the period of strong overlap the modulation of the DC current occurs for delays on the sub-fs time scale resulting from the strongly varying maximum instantaneous laser intensity as a function of pump-probe delay (Fig.~\ref{fig4}b). Assuming that the charge transfer is governed by the central peak of the combined laser pulse, a simple estimate in analogy to Eq.~\ref{eq:jpge} predicts
\begin{equation} \label{eq:model}
  D(\Delta t) = \mathrm{sgn}(I_c-I_\mathrm{max}(\Delta t)) \, \, \beta_\mathrm{NL} \, \,  I_\mathrm{max}(\Delta t)^{2.8} \quad ,
\end{equation}
where $\mathrm{sgn}$ denotes the sign function and $I_\mathrm{max}(\Delta t)$ is the maximum instantaneous laser intensity for pump-probe delay $\Delta t$ (Fig.~\ref{fig4}b). In Eq.~\ref{eq:model}, we denote the non-linear generalization of the photogalvanic tensor by $\beta_\mathrm{NL}$. This simple model reproduces the temporal variation of $D(\Delta t)$ in the full TD-DFT calculations remarkably well, underlining that the maximum instantaneous laser field drives the non-linear photogalvanic effect through tunneling near the field maximum. 

In conclusion, we predict a non-linear extension of the photogalvanic effect into the strong-field regime giving rise to ultrafast DC currents in insulators illuminated by multi-femtosecond laser pulses. We observe a strongly non-linear intensity dependence and even a reversal of the induced currents above a critical intensity $I_c$ associated with the transition from non-linear polarization currents to tunneling excitation. The charge transfer is rather insensitive to details of the laser pulse shape and carrier-envelope phase but strongly dependent on the maximum instantaneous field strength. The latter may be controlled by the pump-probe delay in a two-pulse setup giving rise to a distinct sign change in the transferred charge as function of the pump-probe delay. 
The non-linear photogalvanic effect opens up opportunities for light-field controlled femtosecond charge separation with relatively modest requirements on the driving laser. Even many-cycle pulses without CEP stabilization can be used as the lattice structure instead of the CEP is employed to break the inversion symmetry along the laser polarization axis. The non-linear photogalvanic effect is conceptually simpler than the CEP dependent charge transfer since no elaborate steering of the conduction band electrons is necessary. Therefore, the effect is robust against changes in the laser pulse parameters. 
This may be advantageous in particular for optical interconnects based on surface plasmon propagation \cite{krausz_attosecond_2014} where the pulse shape and duration of a surface plasmon wave packet is difficult to control. The sharp threshold intensity $I_c$ for reversal of the current may provide a simple route towards femtosecond current switching and, moreover, a sensitive intensity calibration for laser pulses that directly measures the maximum electric field strength in the material. Finally, the photogalvanic effect may also be investigated by associated terahertz emission \cite{gildenburg_optical--thz_2007,*sames_all-optical_2009,*silaev_residual-current_2009}.

This work was supported by the FWF (Austria), SFB-041 ViCoM, SFB-049 Next Lite, doctoral college W1243, and P21141-N16. G.W.\ thanks the IMPRS-APS for financial support. X.-M.T.\ was supported by a Grant-in-Aid for Scientific Research (C24540421) from the JSPS. K.Y.\ acknowledges support by the Grants-in-Aid for Scientific Research Nos.\ 23340113 and 25104702. Calculations were performed using the Vienna Scientific Cluster (VSC) and the supercomputer at the Institute of Solid State Physics, University of Tokyo.

\section*{References}

\bibliography{gw}   

%


\end{document}